\documentclass[10pt,twocolumn,letterpaper]{article}

\usepackage[pagenumbers]{cvpr} 

\usepackage{graphicx}
\usepackage{amsmath}
\usepackage{amssymb}

\usepackage{xcolor}
\usepackage{enumitem}
\usepackage{xspace}
\usepackage{booktabs}
\usepackage{colortbl}
\usepackage{multirow}
\usepackage{makecell}
\usepackage{lipsum} 
\usepackage{gensymb} 

\usepackage[labelsep=period]{caption}
\renewcommand{\paragraph}[1]{\vspace{0.2em}\noindent \textbf{#1 \hspace{0.2em}}}

\definecolor{MyDarkRed}{rgb}{0.46, 0.16, 0.16}
\definecolor{MyDarkBlue}{rgb}{0.16, 0.16, 0.66}

\def\pipelineName{Hi3DGen\xspace}
\def\datasetName{DetailVerse\xspace}
\def\ItoNmethodName{NiRNE\xspace}
\def\NtoGmethodName{NoRLD\xspace}

\definecolor{iccvblue}{rgb}{0.21,0.49,0.74}
\usepackage[pagebackref,breaklinks,colorlinks,allcolors=iccvblue]{hyperref}

\usepackage[marginal]{footmisc}

\newcommand{\wlink}[1]{\textcolor{magenta}{{#1}}}

\def\paperID{6466} 
\def\confName{ICCV}
\def\confYear{2025}

\title{Hi3DGen: High-fidelity 3D Geometry Generation\\from Images via Normal Bridging}

\author{Chongjie Ye$^{1,2}\footnotemark[1]$ \quad Yushuang Wu$^{2}\footnotemark[1]$ \quad Ziteng Lu$^{1}$ \quad Jiahao Chang$^{1}$\\Xiaoyang Guo$^{2}$  \quad Jiaqing Zhou$^{2}$ \quad Hao Zhao$^{3}$ \quad Xiaoguang Han$^{1}\footnotemark[2]$ \vspace{0.3em} \\
{\normalsize $^1$The Chinese University of Hong Kong, Shenzhen} \quad{\normalsize $^2$ByteDance} \quad {\normalsize $^3$Tsinghua University}
}

\begin{document}

\twocolumn[{%
\renewcommand\twocolumn[1][]{#1}%
\maketitle
\begin{center}
    \centering
    \captionsetup{type=figure}
    \vspace{-8mm}
    \includegraphics[width=0.94\textwidth]{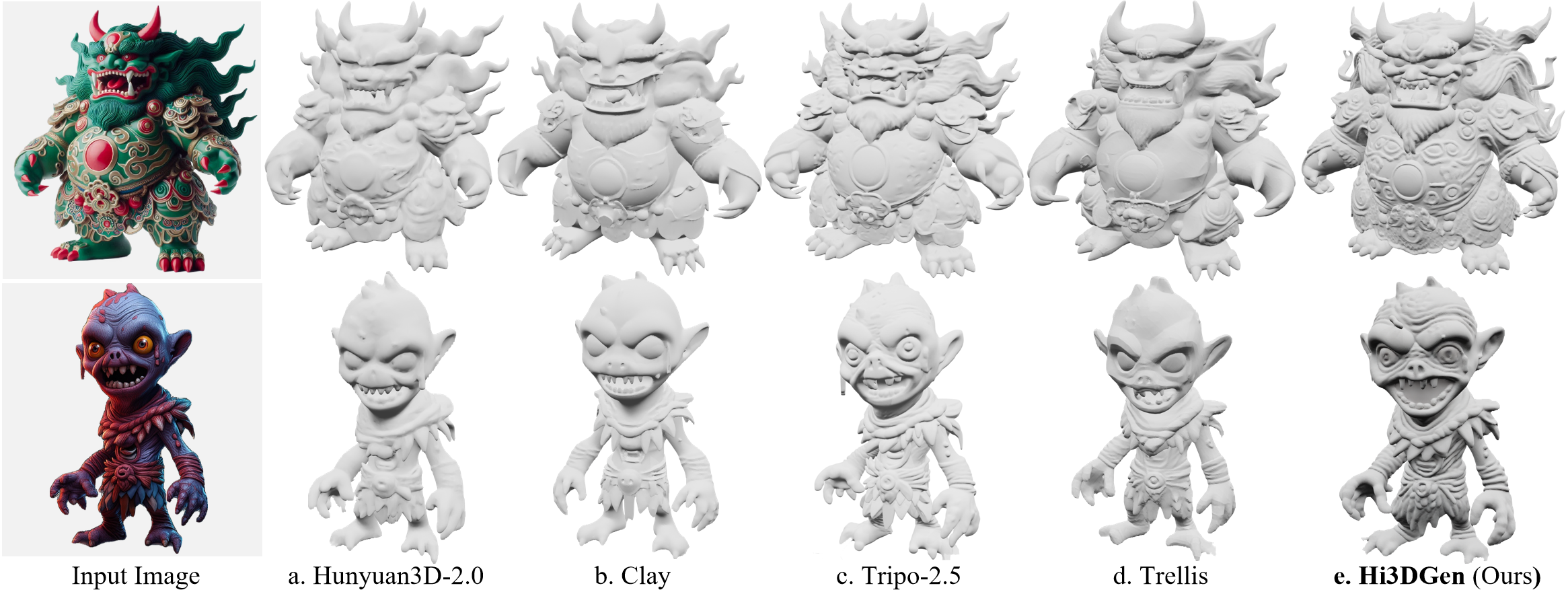}
    \vspace{-1mm}
    \label{fig:teaser}
\end{center}%
}]

\footnotetext[1]{Equal Contribution.}
\footnotetext[2]{Corresponding author:
\wlink{hanxiaoguang@cuhk.edu.cn}.}

\begin{abstract}
    With the growing demand for high-fidelity 3D models from 2D images, existing methods still face significant challenges in accurately reproducing fine-grained geometric details due to limitations in domain gaps and inherent ambiguities in RGB images. To address these issues, we propose \textbf{Hi3DGen}, a novel framework for generating high-fidelity 3D geometry from images via normal bridging. Hi3DGen consists of three key components: (1) an image-to-normal estimator that decouples the low-high frequency image pattern with noise injection and dual-stream training to achieve generalizable, stable, and sharp estimation; (2) a normal-to-geometry learning approach that uses normal-regularized latent diffusion learning to enhance 3D geometry generation fidelity; and (3) a 3D data synthesis pipeline that constructs a high-quality dataset to support training. Extensive experiments demonstrate the effectiveness and superiority of our framework in generating rich geometric details, outperforming state-of-the-art methods in terms of fidelity. Our work provides a new direction for high-fidelity 3D geometry generation from images by leveraging normal maps as an intermediate representation.

\end{abstract}

\section{Introduction}
\label{sec:intro}

With the rapid advancement of computer vision and graphics technologies, the task of generating 3D models from 2D images has garnered significant attention in both academic and industrial domains. 
Despite significant advancements in recent years, existing methods remain inadequate in generating 3D models that sufficiently reflect the geometric details present in the input images, especially when dealing with real-world input images, which typically exhibit complex and rich geometric characteristics. Nevertheless, the ability to faithfully reproduce these geometric details in 3D generations is of paramount importance, as it directly influences the models' realism, precision, and overall applicability in practical scenarios.


Current state-of-the-art techniques for 3D generation from 2D images often rely on deep learning models to learn the direct mapping from the 2D RGB image to the 3D geometry. While these methods have shown promising results~\cite{li2024craftsman, wu2024direct3d, zhang2024clay}, their ability in producing fine-grained geometric details is inherently limited by several key factors. First, the scarcity of high-quality 3D training data restricts the model's ability to learn detailed geometric features. Second, there exists a significant domain gap between the training images (often rendered from synthetic 3D meshes) and test images of various possible styles, leading to suboptimal performance in practical applications. Third, the inherent ambiguity in RGB images, caused by lighting, shading,  or complex object textures, further complicates the extraction of fine-grained geometric information.

To address these limitations, we propose to leverage normal maps as an intermediate representation to bridge the mapping from 2D RGB images to 3D geometry. Normal maps, which encode surface orientation information, offer several advantages for this task. First, by introducing strong 2D priors to process RGB images into normal maps, we can effectively alleviate the domain gap between synthetic training data and real-world applications, which eases the 2D-to-3D mapping learning. Second, normal maps, as a 2.5D representation, provide clearer geometric cues compared to RGB images, thereby having the potential of guiding the geometry learning more effectively, especially in producing fine-grained geometric details.

In this paper, we introduce \textbf{\pipelineName}, a novel framework for high-fidelity 3D geometry generation from images via normal bridging. The framework consists of three key components: (i) an image-to-normal estimator (\ItoNmethodName) that achieves generalizable, stable, and sharp normal estimation through a noise-injected regressive network with dual-stream training to decouple the representation learning of low- and high-frequency image patterns; (ii) a normal-to-geometry learning approach (\NtoGmethodName) that employs normal-regularized latent diffusion learning to provide explicit 3D geometry supervision during training, significantly enhancing generation fidelity; and (iii) a 3D data synthesis pipeline that constructs the \datasetName dataset, containing high-quality synthesized 3D assets, serving as important complementary of humman-created ones, to support the training of \ItoNmethodName and \NtoGmethodName. 
Our framework generates rich, fine-grained geometric details, surpassing state-of-the-art (SOTA) approaches in terms of generation fidelity, as shown in teaser figure.

\paragraph{Contributions} Our key contributions are as follows:  
\begin{itemize}[itemsep=0pt,parsep=0pt,topsep=2bp]
    \item We propose Hi3DGen, the first framework that leverages normal maps as an intermediate representation to bridge the gap between 2D images and 3D geometry, addressing the limitations of existing methods in generating fine-grained details;
    \item We introduce \ItoNmethodName, which decouples the low-high frequency learning with noise-injected dual-stream training to achieve robust, stable, and sharp normal estimation from input images;
    \item We develop a data synthesis pipeline and construct the \datasetName dataset, which contains high-quality synthesized 3D assets to support the training of our framework. We will also release this dataset and hope it can inspire related research;
\end{itemize}

\section{Related Work}
\label{sec:related_works}

\paragraph{Datasets for 3D Generation} 
Early 3D datasets typically encompass small-scale objects from a limited category range~\cite{chang2015shapenet, wu2015modelnet, collins2022abo}. To address this limitation, researchers endeavor to expand 3D data repositories through scanning or multi-view photography~\cite{jampani2023navi, downs2022gso, wu2023omniobject3d, uy2019scanobjectnn, reizenstein2021co3d}. This approach leads to the creation of large-scale datasets such as MVImgNet~\cite{yu2023mvimgnet, wu2024mvimgnet2}. However, the quality of the constructed data often falls short of the requirements for direct application in 3D generation tasks.
Recently, larger-scale datasets have been constructed by aggregating available human-created 3D assets from a wide range of online sources~\cite{deitke2023objaverse, deitke2024objaversexl}. However, among the 10 million 3D assets in Objaverse-XL~\cite{deitke2024objaversexl}, 5.5 million are from GitHub~\cite{github}, raising license concerns and high quality variety necessitating costly data cleaning, and another 3.5 million from Thingiverse~\cite{thingiverse} lack textures required by existing 3D generation pipelines. The remaining objects, mainly from Objaverse-1.0~\cite{deitke2023objaverse}, exhibit a severe imbalance, characterized by a scarcity of high-quality assets with complex geometric structures and rich surface details.
This imbalance is a common issue in datasets of human-created 3D meshes, resulting in networks generating simplistic 3D models with significant loss of detail. To address this gap, this paper explores synthesizing 3D data with high semantic variety, geometric structure diversity, and surface detail richness, and utilizes them in the context of 3D generation as a non-trivial complement to human-created 3D assets.

\begin{figure*}[tb] \centering
    \includegraphics[width=\textwidth, height=0.4\textwidth]{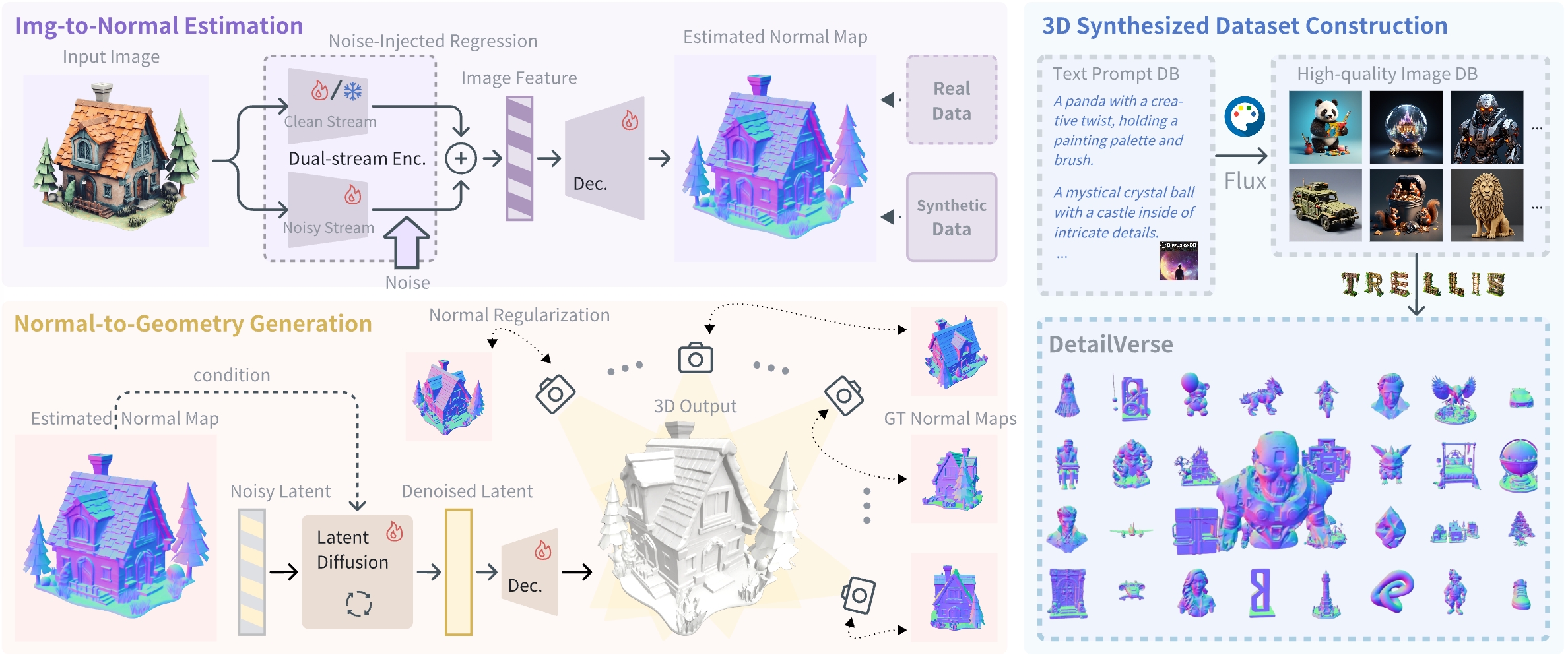}
    \caption{Overview of the proposed normal-bridged 3D geometry generation method. Our \pipelineName comprises three components: an image-to-normal estimator, a normal-to-geometry generator, and a synthesized dataset (\datasetName) construction pipeline.}
    \label{fig:overview}
    \vspace{-3mm}
\end{figure*}


\paragraph{Normal Estimation} 
Monocular methods can be primarily divided into diffusion-based and regression-based approaches. 
Regression-based methods have advanced from early handcrafted features~\cite{Hoiem_Efros_Hebert_2005, Hoiem_Efros_Hebert_2007} to deep learning techniques~\cite{eigen2015predicting, Wang_Geraghty_Matzen_Szeliski_Frahm_2020, Zhang_Cui_Xu_Yan_Sebe_Yang_2019}. Recent progress includes leveraging large-scale data~\cite{eftekhar2021omnidata}, estimating per-pixel normal probability distributions~\cite{Bae_Budvytis_Cipolla_2021}, adopting vision transformers~\cite{Ranftl_Bochkovskiy_Koltun_2021}, and conducting inductive bias modeling~\cite{bae2024dsine}. Though conducting deterministic prediction that ensures higher stability, regression-based methods struggle with generating fine-grained sharp details. 
Diffusion-based normal estimation has emerged with the adaptation of powerful text-to-image models~\cite{rombach2021highresolution, zhang2023controlnet, peebles2023scalable}. For instance, Geowizard~\cite{fu2024geowizard} incorporates a geometry switcher to handle diverse data distributions. Considering high-variance results caused by the inherent stochastic nature of diffusion processes~\cite{everaert2024exploiting}, strategies such as affine-invariant ensembling~\cite{ke2023marigold, fu2024geowizard} and one-step generation~\cite{xu2024genpercept} have been explored but come with computational intensity and oversmoothing issues. StableNormal~\cite{ye2024stablenormal} improves estimation stability by reducing diffusion inference variance via a coarse-to-fine strategy, but it remains challenged by imperfect stability. Differently, by deeply exploring the root causes of the sharpness produced by diffusion-based methods, we novelly propose a noise-injected regressive method to enable both sharp and stable estimations, with a dual-stream training strategy to fully utilize training data from different domains.

\paragraph{Normal Maps in 3D Generation} 
Normal maps, which provide detailed geometric cues, have been widely used to enhance the fidelity and consistency of 3D reconstructions~\cite{saito2020pifuhd, yu2022monosdf, wang2022neuris, xiu2022icon, xiu2023econ, wei2024normal, cao2024supernormal, brument2024rnb}. Recently, they have also been explored in 3D generation. SDS-based methods~\cite{poole2022dreamfusion, wang2023sjc} render normal maps alongside RGB images in optimization to regularize geometry~\cite{qiu2024richdreamer, huang2024humannorm, gu2025tetrahedron}. Other works use multi-view diffusion followed by reconstruction or fusion, generating normal images to complement RGB data and improve accuracy~\cite{long2024wonder3d, lu2024direct2, wu2024unique3d, pang2024envision3d, sun2024dreamcraft3d++, bala2024edify3d, yang2025genvdmgeneratingvectordisplacement}. Though suffering from generating smooth surface details due to multi-view inconsistency, they have shown the significant potential of normal maps in enhancing 3D generation.
In parallel, methods conducting 3D native diffusion based on 3D representations such as feature volumes, Triplane~\cite{chan2022eg3d_triplane}, 3D Gaussians~\cite{kerbl20233dgaussian} have leveraged normal maps by decoding them into meshes and applying normal rendering losses to regularize surfaces~\cite{dong2023ag3d, xu2024instantmesh, zheng2024mvd, liu2024meshformer}. However, these approaches often face limitations due to high memory requirements for high-resolution 3D representations. Meanwhile, methods focusing on latent code diffusion have achieved state-of-the-art performance~\cite{zhang2024clay, wu2024direct3d, li2024craftsman, xiang2024trellis, li2025triposg}. However, the use of normal maps in this paradigm remains underexplored, where normal maps can not directly regularize the diffusion learning in the highly abstract latent space. Notable examples include CraftsMan~\cite{li2024craftsman}, which uses normal map refinement as a post-processing step, and Trellis~\cite{xiang2024trellis}, which incorporates normal rendering loss during VAE training.
Our approach uniquely emphasizes the critical role of normal maps in bridging image-to-3D generation and introduces a novel method to effectively integrate normal supervision into the diffusion learning of 3D latent codes, addressing limitations of prior work.








\section{Method}
\label{sec:method}
This section outlines the proposed \pipelineName framework, which aims to bridge the learning of 2D-to-3D lifting with 2.5D representation, normal map. Dividing the image-to-geometry generation into two parts, image-to-normal estimation and normal-to-geometry mapping, our framework consists of a dual-stream normal estimator for prediction stability and sharpness (Sec.~\ref{sec:image2normal}) and an online normal regularizer for fine-grained generation details and image-geometry consistency in diffusion training (Sec.~\ref{sec:normal23d}). We further propose a synthesized 3D dataset, which contains numerous generated 3D data of complex geometry structure and rich surface details, to facilitate sharp normal estimation and detailed 3D geometry generation (Sec.~\ref{sec:dataset}). An overview of the whole framework is visualized in Fig.~\ref{fig:overview}.

\begin{figure}[tb] \centering
    \includegraphics[width=0.48\textwidth]{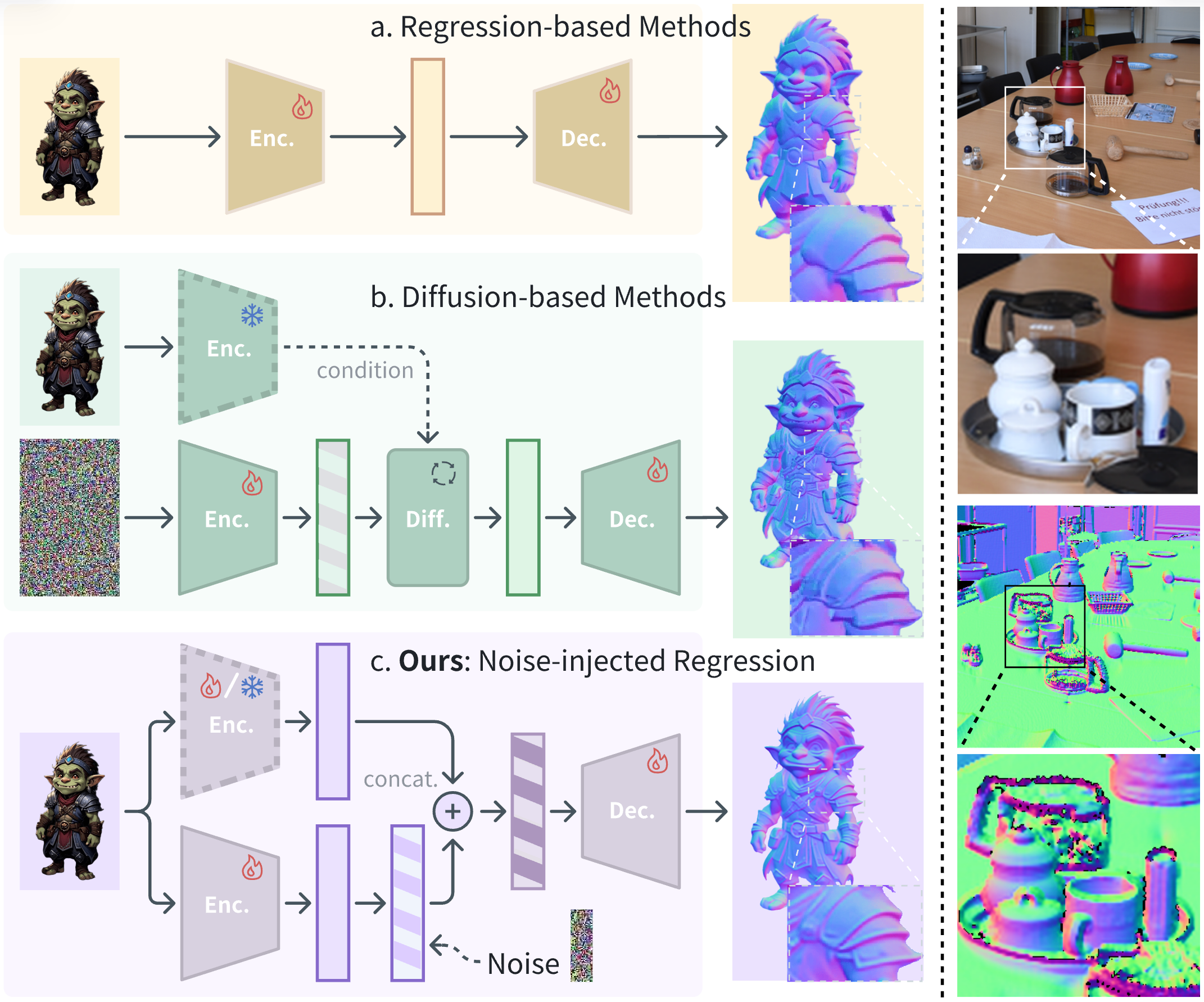}
    \caption{Left part: Illustration of Noise-injected Regressive Normal Estimation; Right part: Noisy label at high-frequency regions in real-domain data.} \label{fig:method_i2n}
    \vspace{-3mm}
\end{figure}

\subsection{Noise-Injected Regressive Normal Estimation}
\label{sec:image2normal}
SOTA monocular normal estimation methods are mainly divided into diffusion-based and regression-based approaches. The former produces sharper results yet suffers from instability and spurious details due to their inherent probabilistic nature, while the latter offers stable one-step predictions but lacks sharpness. We first analyze the reason for sharper estimations of diffusion-based methods from the viewpoint of frequency domain. Then we propose integrating noise injection, the key mechanism in diffusion learning, into a regressive framework to enhance its sensitivity to high-frequency patterns, as illustrated in Fig.~\ref{fig:method_i2n}. Based on this, we further develop a dual-stream architecture to decouple the low- and high-frequency representation learning for both generalizability and sharpness, with a domain-specific training strategy to stimulate the decoupled learning. 

\paragraph{Noise Injection} Considering normal sharpness usually appears at high-frequency image regions like edges and cavities, we begin by analyzing from the frequency domain the underlying mechanisms that enable sharp normal estimation results of diffusion-based methods. Defining the diffusion process with a stochastic differential equation:
\begin{equation}
\begin{aligned}
x_t = x_0 + \int_0^t g(s)dw_t,
\end{aligned}
\end{equation}
where the initial state $X_0$ evolves over time $t\in[0,T]$ to become $x_t$ and $w_t$ is a Wiener process (Brownian motion) representing injected random noise. By conducting Fourier transformation to this process, we can obtain the signal-to-noise ratio (SNR) of any frequency component $\omega$ at timestep $t$:
\begin{equation}
\begin{aligned}
\text{SNR}(\omega, t) = \frac{|\hat{x}_0(\omega)|^2}{\int_0^t |g(s)|^2ds},
\end{aligned}
\end{equation}
which is only subject to $\hat{x}_0(\omega)$ because the power of noise is equal over all $\omega$.
Since natural images exhibit low-pass characteristics, \ie, $|\hat{x}_0(\omega)|^2 \propto |\omega|^{-\alpha}$ where $\alpha > 0$ represents the attenuation coefficient, the high-frequency components in $x_t$ has a faster SNR degradation than low-frequency ones as the diffusion process progresses. This prompts that the model gets a stronger supervision at high-frequency regions in $x_t$, which encourages the model to focus more on capturing and predicting sharp details. Inspired by this, we integrate the noise injection technique into regression-based methods to encourage learning more high-frequency information.

\paragraph{Dual-Stream Architecture} Compared to high-frequency features influencing the prediction sharpness, low-frequency features, conveying more overall structure information~\cite{chen2024bracketing, han2023dual}, are important for the generalizability in low-level vision tasks~\cite{li2022transfering}. To decouple these two kinds of features, we encode the input image through two independent streams: one processes the original image without noise injection to robustly capture low-frequency details (clean stream), while the other processes the noise-injected image to focus on high-frequency details (noisy stream). The latent representations from both streams are concatenated in a ControlNet-style manner~\cite{zhang2023controlnet} and fed into the decoder for final predictions, in a regression manner. This design uses noise injection in one stream to encourage high-frequency representation learning, and also maintains another clean stream to perceive the original image for regression, which effectively integrates the strengths of diffusion-based methods into a regressive method. An illustration of the method is presented in Fig.~\ref{fig:method_i2n}(c).

\paragraph{Domain-Specific Training} To encourage the decoupled representation learning in two streams, we design a domain-specific training strategy to optimize the network by delicately utilizing training data from different domains. Previous methods mix real- and synthetic-domain data in training to enhance the generalizability. However, real-domain data, limited by the collection environment and the precision of scanners, suffer from noisy labels especially at object edges (see a visualized example in Fig.~\ref{fig:method_i2n} right part), which hinder accurate learning at high-frequency details. In contrast, synthetic domain data, constructed via rendering from 3D ground truth, can provide precise high-frequency labels, while it is limited by the domain gap with real images in application.  
Therefore, we first train the network using real-domain data to capture low-frequency information for strong generalizability. In the second stage, we fine-tune the noisy stream using synthetic-domain data while freezing the parameters of the other stream. This allows the noisy stream to focus on learning high-frequency details as a residual component of outputs by the clean stream. The domain-specific training not only well utilizes the training data from real and synthetic domains according to their strengths, but also properly encourages the optimization of dual streams for decoupled representation learning. 

\begin{figure}[tb] \centering
    \includegraphics[width=0.48\textwidth]{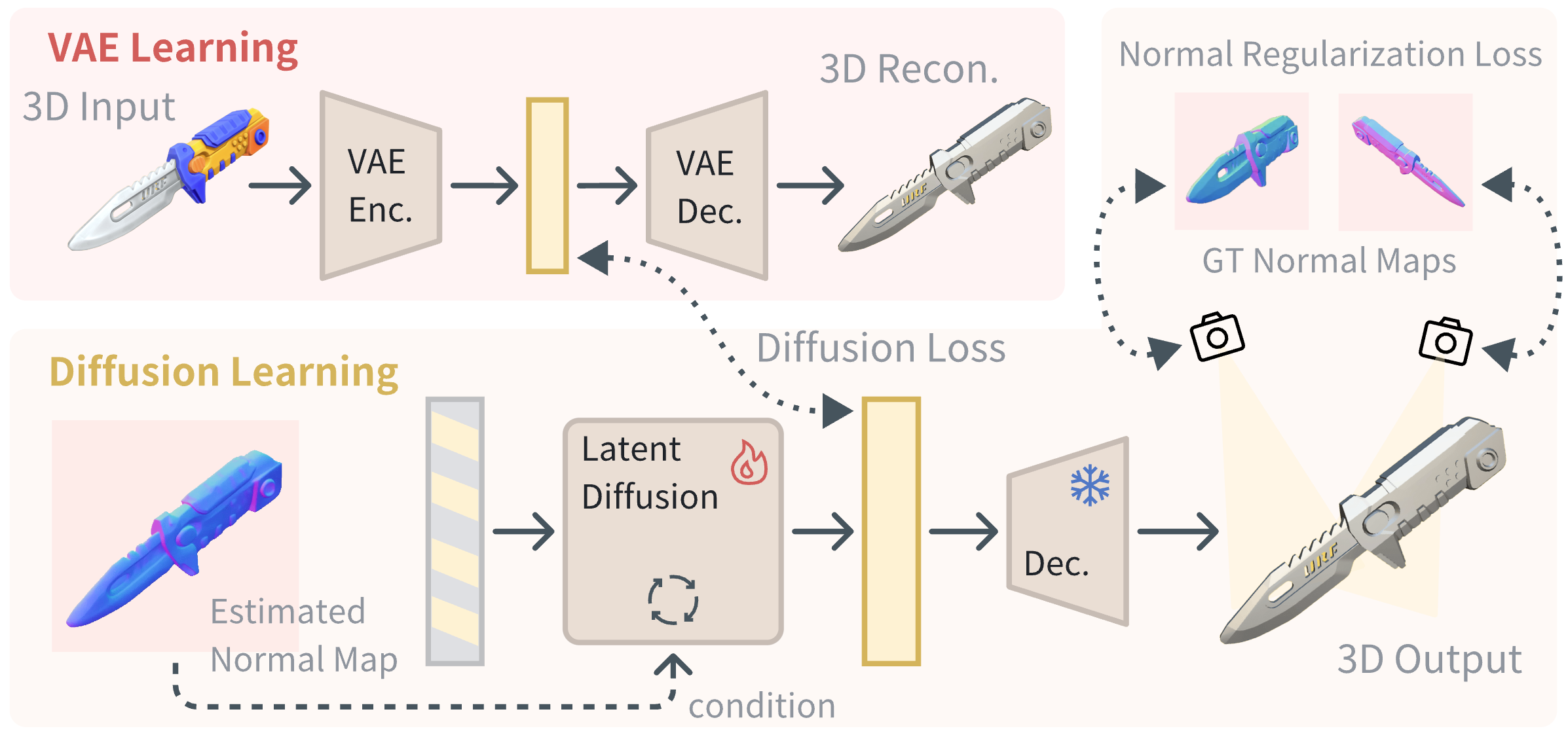}
    \caption{An illustration of Normal-Regularized Latent Diffusion.}
    \label{fig:method_n2g}
    \vspace{-3mm}
\end{figure}

\subsection{Normal-Regularized Latent Diffusion}
\label{sec:normal23d}
State-of-the-art 2D-to-3D generation methods rely on 3D latent diffusion, which represents 3D geometries in a compact latent space so that the 2D-to-3D mapping can be learned more efficiently~\cite{zhang2024clay, wu2024direct3d, li2024craftsman, xiang2024trellis}. However, these methods suffer from easy loss of details or detail-level inconsistency with the input images (see examples in Fig.~\ref{fig:teaser}). Except for the ambiguity of RGB image input in indicating fine-grained geometries, another important reason is the indirect supervision from the latent space only, where the geometry information, especially fine-grained details, is usually greatly compressed to ensure compactness for diffusion learning.  

\paragraph{Latent Diffusion} We first formulate the typical latent diffusion process in 3D generation methods. A Variational Auto-Encoder (VAE) is trained to encode any 3D geometry $X$ into a latent representation $x_0$ and decode it back to the original geometry $\hat{X}$:
\begin{equation}
\begin{aligned}
x_0 = E(X), ~~~\hat{X} = D(x_0),
\end{aligned}
\end{equation}
where $E(\cdot)$ and $D(\cdot)$ denote the encoder and decoder, respectively. The reparameterization process is omitted for simplicity.
The image-conditioned diffusion process constructs $x_t$ by injecting noise into $x_0$ at a given timestep $t$ and learns to recover $x_0$ from $x_t$. Flow matching is commonly used to address it, which aims to learn a continuous transformation by modeling the time-dependent velocity field, with the loss function formulated as:
\begin{equation}
\begin{aligned}
\mathcal{L}_\text{LDM} =  \mathbb{E}_{t, x_0, x_t}\Big[ \big\|\mathbf{v}_\theta(x_t, t) - \mathbf{u}(x_t, t)\big\|^2\Big],
\end{aligned}
\end{equation}
where $\theta$ denotes the network parameters, $\mathbf{u}(x_t, t) = \nabla_{x_t} \log p(x_t|x_0)$ is the true velocity field, and the image/text condition is implicitly included. 

\paragraph{Normal Regularization} Regularization in the 3D geometry space allows for more precise supervision, especially over surface details. Therefore, we propose an enhanced loss function with explicit normal map regularization:
\begin{equation}
\begin{aligned}
\mathcal{L}_\text{Norld} = \mathcal{L}_\text{LDM} + \lambda \cdot \mathcal{R}_{\text{Normal}}(\hat{x}_0),
\end{aligned}
\end{equation}
where $\hat{x}_0$ represents the predicted clean sample and $\mathcal{R}_{\text{normal}}$ is the proposed regularization term:
\begin{equation}
\begin{aligned}
\mathcal{R}_{\text{Normal}}(\hat{x}_0) = \mathbb{E}_{v}\Big[\big\|R_v(D(\hat{x}_0)) - N_v\big\|^2\Big],
\end{aligned}
\end{equation}
where the predicted target latent $\hat{x}_0$ is decoded into explicit 3D geometry, $R_v$ renders the normal map from viewpoint $v$, and $N_v$ denotes the corresponding ground truth normal map. Note that this regularization is conducted online during the diffusion training process, as shown in Fig.~\ref{fig:method_n2g}, rather than in a post-processing stage. This actively guides the training of diffusion networks and aligns the predicted latent with a distribution that contains rich details consistent with the input images. 

\begin{figure}[tb] 
    \centering
\includegraphics[width=0.48\textwidth,height=0.3\textwidth]{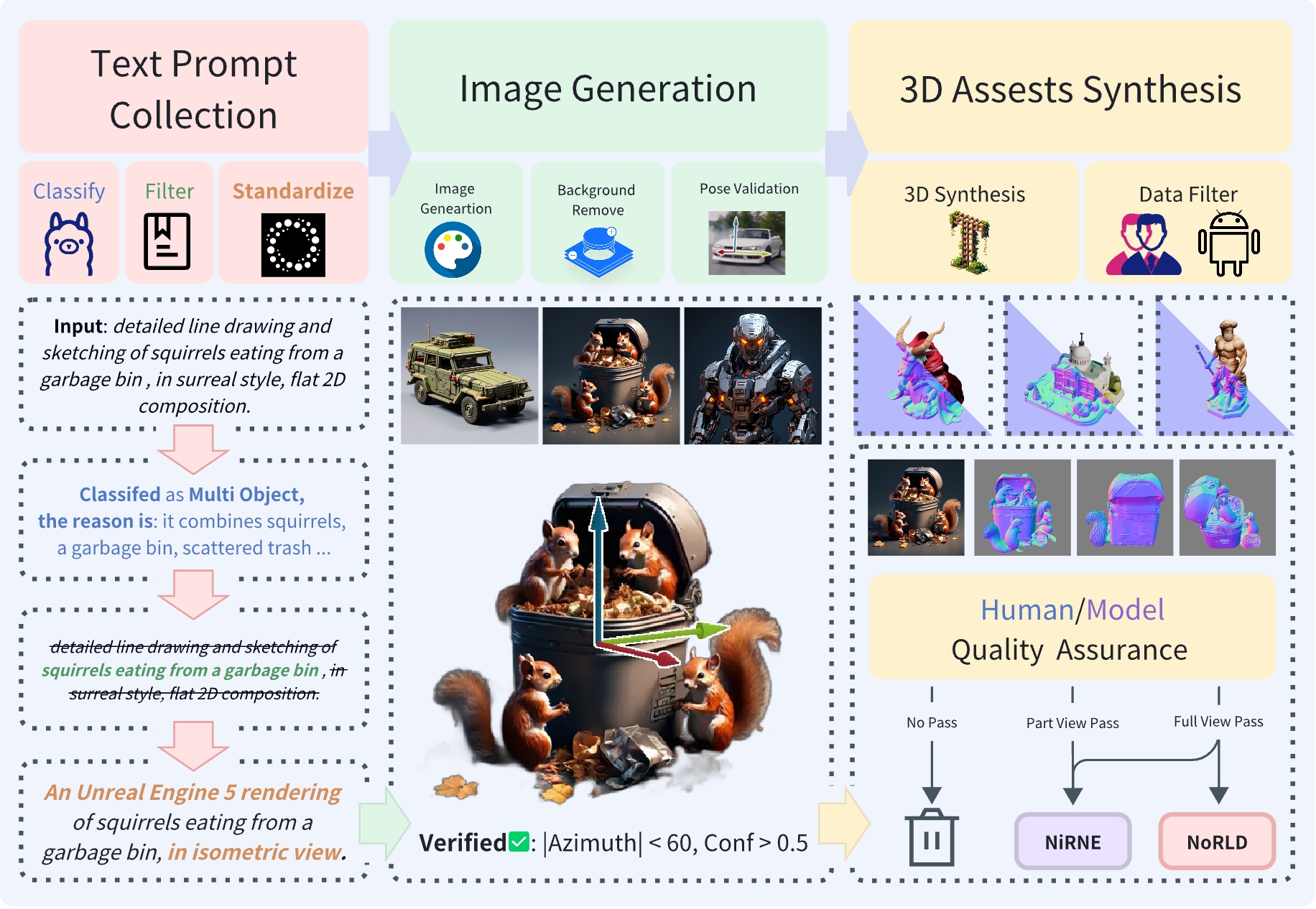}
    \caption{The procedure of \datasetName Construction.}
    \label{fig:method_dv}
    \vspace{-5mm}
\end{figure}

\subsection{\datasetName Dataset}
\label{sec:dataset}
High-quality 3D data is essential in the training of our \ItoNmethodName to provide clean normal labels and \NtoGmethodName for high-fidelity 3D generation. Although Objaverse~\cite{deitke2023objaverse, deitke2024objaversexl} provides a substantial number of image-normal and normal-geometry training pairs, the majority of the 3D assets exhibit simple structures and plain surface details, as shown in Tab.~\ref{tab:dataset_comparison}. This limitation restricts the generation capabilities of \pipelineName. Given the prohibitive cost of manually creating high-quality 3D assets, we propose a 3D data synthesis pipeline to perform Text$\xrightarrow{}$Image$\xrightarrow{}$3D generation.
By using advanced generators integrated with meticulous prompt engineering and data cleaning, this pipeline leads to a dataset \datasetName of 700k synthesized 3D assets with considerably complex structures and rich details. 

\paragraph{Dataset Construction} We initiate the 3D data synthesis process with text prompts rather than image prompts because text prompts allow for more straightforward control of semantic diversity, thereby ensuring the variety of final geometries. we first sourced approximately 14M high-quality raw prompts from DiffusionDB~\cite{wang2023diffusiondb}. A LLaMA-3-8B model~\cite{touvron2023llama} is adopted for classification to filter out complex scenes. Then, a rule-based filtering method to eliminate stylistic modifiers, together with a  LLaMA-3-13B~\cite{touvron2023llama} for structural standardization to ensure consistent formatting. In the second step, we employ the SOTA Flux.1-Dev~\cite{flux} for text-to-image generation. Additionally, we specify the text prompt condition to control the viewpoint and lightning in generation, and conduct pose validation using OrientAnything~\cite{wang2024orient} to filter ones with special viewpoints, which is important to guarantee stable 3D generation. In the third step We employ Trellis~\cite{xiang2024trellis}, a SOTA 3D generator, to conduct image-to-3D synthesis. Finally, a rigorous data cleaning process that combines expert evaluation with automated assessment preserves 700k high-quality meshes. 

\paragraph{Dataset Statistics}
We present the model number in the dataset and the mean sharp edge number in each model in Tab.~\ref{tab:dataset_comparison} to show the scale and geometric detail richness of our \datasetName dataset. The sharp edge detection follows the implementation in Dora-Bench~\cite{chen2024dora}. 
The synthesized assets in \datasetName present rich surface details, as presented by the examples in the blue block of Fig.~\ref{fig:overview}.
\begin{table}[tb]
\centering
\footnotesize
\setlength{\tabcolsep}{5pt}
\renewcommand{\arraystretch}{1.3}
\caption{Comparison of 3D object dataset statistics. The numbers X/Y in the third column means the Mean/Medium number.  }
\begin{tabular}{
p{1.9cm}>
{\centering\arraybackslash}p{1.2cm}>
{\centering\arraybackslash}p{1.8cm}>
{\centering\arraybackslash}p{1.8cm}>
{\centering\arraybackslash}p{1.2cm}}
\hline
\rowcolor[gray]{0.97} 
\textbf{Dataset} & \textbf{Obj \#} & \textbf{Sharp Edge \#}& \textbf{Source} \\
\hline

GSO~\cite{downs2022gso} & 1K & 3,071 / 1,529  & Scanning \\
Meta~\cite{siddiqui2024meta} & 8K & 10,603 / 6,415  & Scanning \\
ABO~\cite{collins2022abo} & 8K & 2,989 / 1,035  & Artists \\
3DFuture~\cite{fu20213dfuture} & 16K & 1,776 / 865~~  & Artists \\
HSSD~\cite{khanna2024hssd} & 6K & 5,752 / 2,111  & Artists \\
ObjV-1.0~\cite{deitke2023objaverse} & 800K & 1,520 / 452  & Mixed \\
ObjV-XL~\cite{deitke2024objaversexl} & 10.2M & 1,119 / 355 & Mixed \\
\rowcolor[gray]{0.92} \textbf{\datasetName} & 700K & \textbf{45,773 / 14,521}  & Synthesis  \\
\hline
\end{tabular}
\vspace{-3mm}
\label{tab:dataset_comparison}
\end{table}

\section{Experiments}
\label{sec:Experiments}

\subsection{Experiment Setup}
\label{sub:exp_setup}

\paragraph{Dataset} For image-to-normal training, we utilize two complementary datasets. One is a diverse realistic dataset following Depth-pro\cite{bochkovskii2024depthpro}. Another contains synthetic data consisting of 20M RGB-to-normal pairs created by rendering 40 images per asset from 500k \datasetName assets. For normal-to-geometry training, we curate a large-scale dataset comprising 170K cleaned 3D assets from Objaverse~\cite{deitke2023objaverse} and 700K synthesized 3D assets from our \datasetName. We render 40 images per asset following Trellis~\cite{xiang2024trellis}. 
For evaluation, the generalization ability of the image-to-normal estimator in real scenes is validated on the the reconstruction dataset LUCES-MV~\cite{logothetis2024lucesmvmultiviewdatasetnearfield}. All images for visual comparison and user studies are collected from Hyper3D website~\cite{rodin}, Hunyuan3D-2.0 project page~\cite{hunyuan3d2}, and Dora project page~\cite{dora}. 

\paragraph{Implementation Details} 
For image-to-normal, we adopt GenPercept~\cite{xu2024genpercept} architecture for Normal Regression network. We initialize the encoder and decoder weights from the Stable Diffusion V2.1~\cite{rombach2022high}, finetuned using the AdamW optimizer with a fixed learning rate of $3 \times 10^{-5}$. For normal-to-geometry, we build upon the Trellis~\cite{xiang2024trellis}, incorporating classifier-free guidance (CFG)~\cite{ho2021cfg} with a drop rate of 0.1 and AdamW~\cite{loshchilov2017decoupled} optimizer with a fixed learning rate of $1 \times 10^{-4}$. For the normal-to-geometry training stage, we finetune the Large variant of Trellis using 8 NVIDIA A800 GPUs (80GB each) for 50k steps with a batch size of 256. During inference, we set the CFG strength to 3.0 and use 50 sampling steps to achieve optimal results.

\paragraph{Evaluation Metrics} 
For the evaluation of image-to-normal estimation, we basically use normal angle error (NE) to measure the overall prediction accuracy, measured in degrees. We additionally use the metric Sharp Normal Error (SNE) following Dora~\cite{chen2024dora} to give emphasis on sharp edges where geometric details are most salient. For evaluating normal-to-geometry conversion, we render normal maps from 22  viewpoints around each object, which is used for compute NE and SNE to measure the overall and detailed geometry accuracy, respectively. More implementation details are included in the supplementary details. 



\paragraph{Competitive methods}
We compare our \textit{\ItoNmethodName} with SOTA normal estimators across different methodological categories. The comparison includes regression-based methods (Lotus~\cite{he2024lotus} and GenPercept~\cite{xu2024genpercept}), diffusion-based approaches (GeoWizard~\cite{fu2024geowizard} and StableNormal~\cite{ye2024stablenormal}). 
Besides, \textit{\pipelineName} is compared with existing SOTA 3D generation methods including open-sourced CraftsMan-1.5~\cite{li2024craftsman}, Hunyuan3D-2.0~\cite{zhao2025hunyuan3d2}, Trellis~\cite{xiang2024trellis}, and close-sourced Clay\cite{zhang2024clay}, Tripo-2.5~\cite{tripo}, and Dora~\cite{chen2024dora}. Note that Dora has not released its testing API, so we compare with Dora using the examples on its project page.

\begin{figure}[tb] \centering
    \includegraphics[width=0.48\textwidth, height=0.2\textwidth]{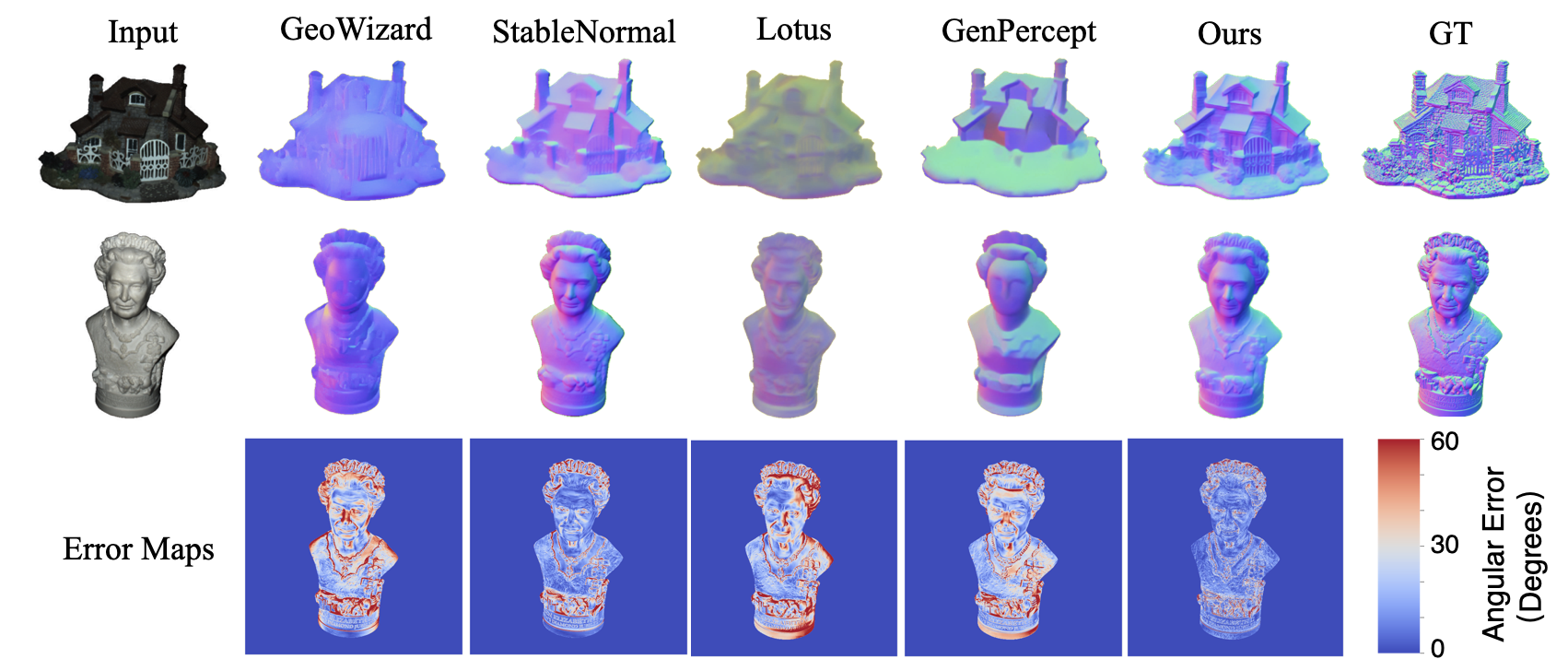}
    \caption{Normal estimation results comparison.}
    \label{fig:result_i2n}
    \vspace{-3mm}
\end{figure}

\subsection{Image-to-Normal Estimation}
\label{sub:res_i2n}
\paragraph{Quantitative Results} We provide the quantitative comparison between our \ItoNmethodName on LUCES-MV and other methods in Tab.~\ref{tab:result_i2n}. It validates that \ItoNmethodName gets significantly superior normal estimation performance to other regression- or diffusion-based methods, in both overall normal accuracy and sharp-region normal accuracy. 

\begin{figure}[tb] \centering
    \includegraphics[width=0.48\textwidth, height=0.19\textwidth]{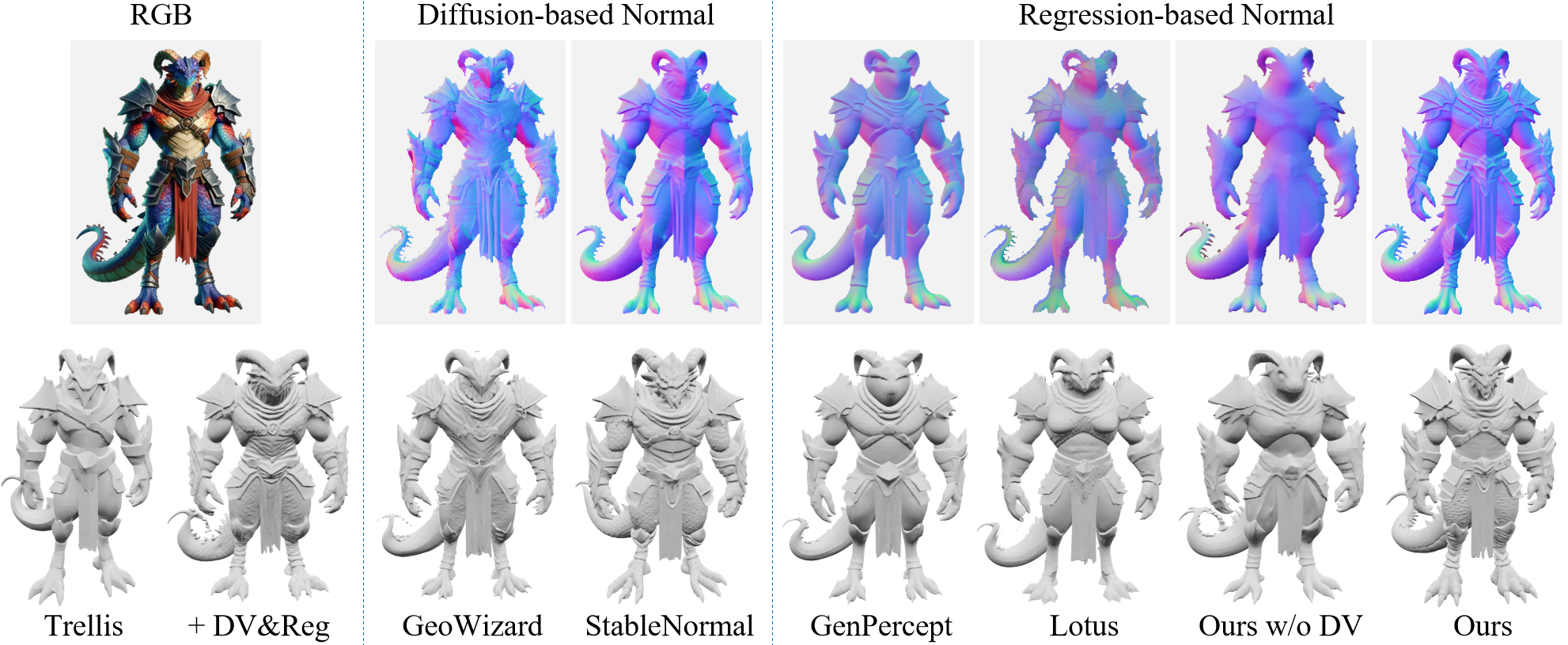}
    \caption{Ablations on the importance of normal bridging.}
    \label{fig:abl_normalbridge}
    \vspace{-3mm}
\end{figure}

\begin{figure}[tb] \centering
    \includegraphics[width=0.48\textwidth, height=0.36\textwidth]{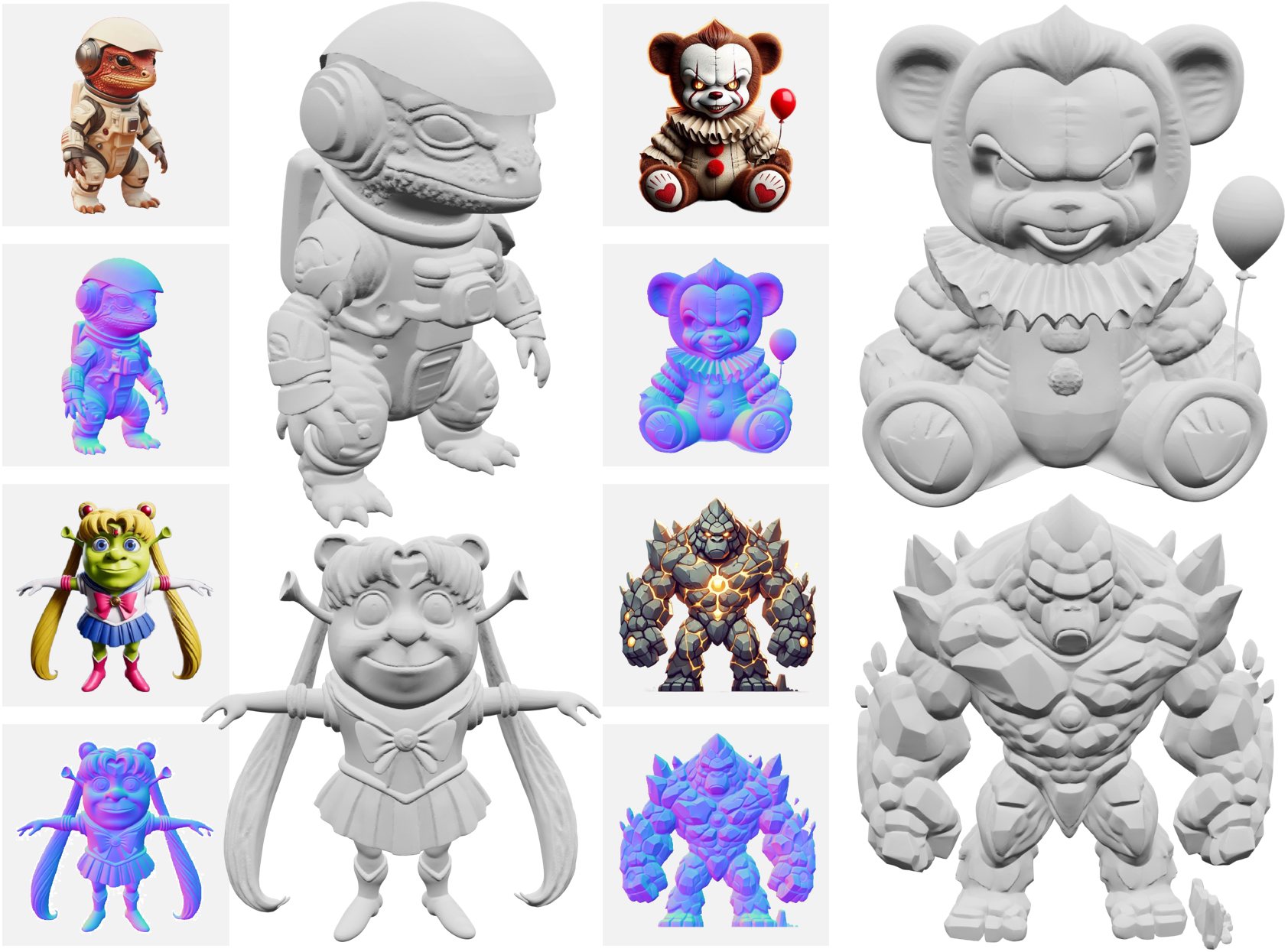}
    \caption{High-fidelity 3D results generated by our \pipelineName.}
    \label{fig:more_our_results}
    \vspace{-3mm}
\end{figure}

\begin{table}
    \centering
    \footnotesize
    \renewcommand{\arraystretch}{1.3}
    \caption{Performance comparison on image normal estimation. We use (Diff.) and (Regr.) to indicate diffusion- and regression-based methods, respectively. \textbf{Bold} indicates best results.}
    \begin{tabular}{
    p{3.7cm}>
    {\centering\arraybackslash}p{1.6cm}>
    {\centering\arraybackslash}p{1.6cm}
    }
         \toprule
         \rowcolor{gray!20} \textbf{Method} & \textbf{NE} $\downarrow$ & \textbf{SNE} $\downarrow$ \\ 
         \hline
         (Diff.) GeoWizard~\cite{fu2024geowizard}& 31.381 & 36.642  \\
         (Diff.) StableNormal~\cite{ye2024stablenormal} & 31.265  & 37.045 \\
         \hline
         (Regr.) Lotus~\cite{he2024lotus} & 53.051 & 52.843 \\
         (Regr.) GenPercept~\cite{xu2024genpercept} & 28.050 & 35.289  \\
         \rowcolor{gray!10} (Regr.) \textbf{\ItoNmethodName} (Ours) & \textbf{21.837}  & \textbf{26.628} \\
         \bottomrule
    \end{tabular}
    \vspace{-5mm}
    \label{tab:result_i2n}
\end{table}

\paragraph{Qualitative Results} A qualitative results is presented in Fig.~\ref{fig:result_i2n}, which shows that our \ItoNmethodName achieves superior estimation performance in (i) robustness with strong generalizability on human and object inputs; (ii) stability with less wrong details than diffusion-based methods (see error maps); and (iii) sharpness especially when compared with regression-based methods. These results further support our related claims in Sec.~\ref{sec:image2normal}. 

\subsection{Normal-to-Geometry Generation}
\label{sub:res_n2g}

\begin{figure*}[tb] \centering
    \includegraphics[width=1.\textwidth]{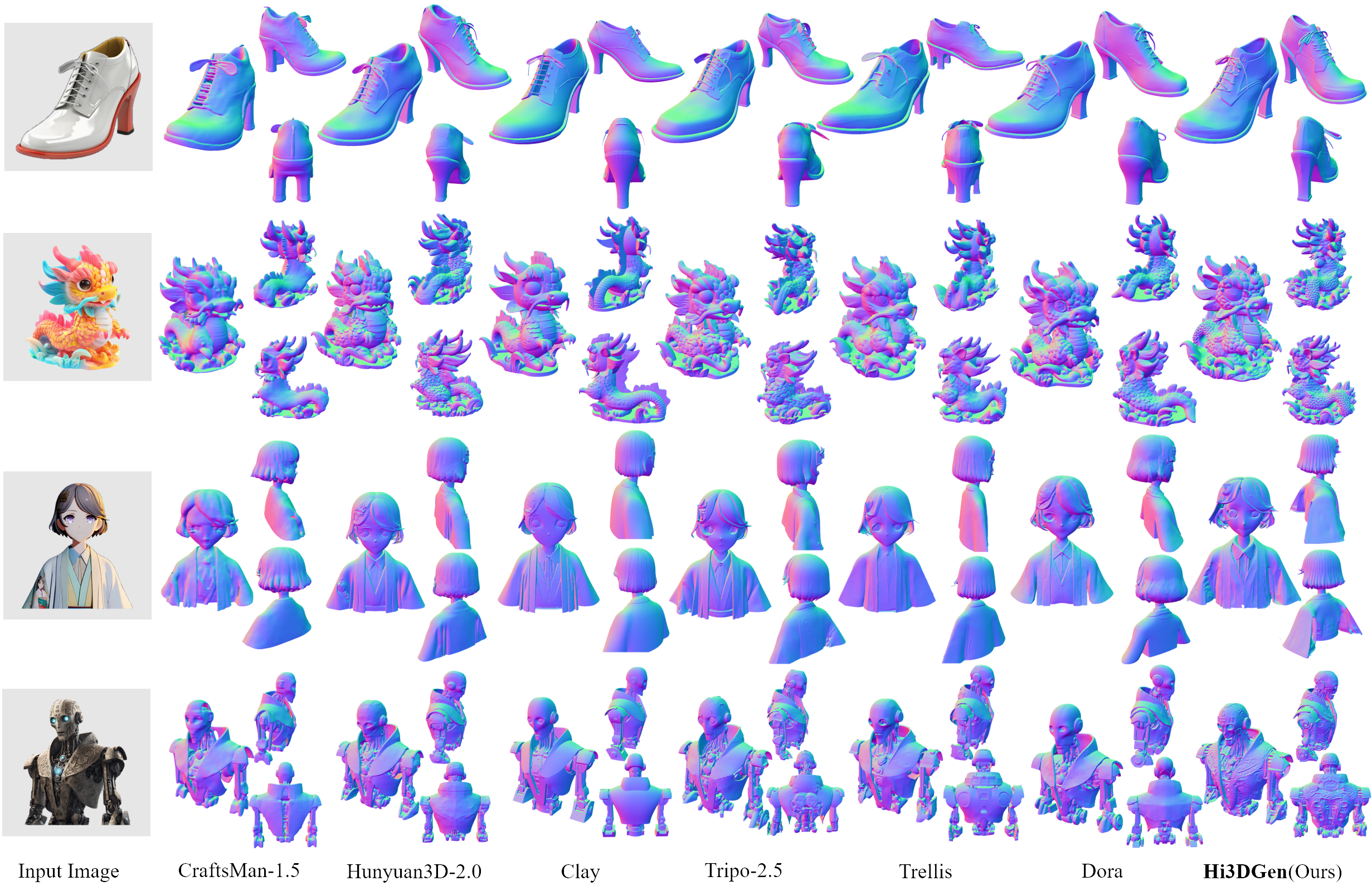}
    \caption{Qualitative 3D generation comparison on samples from Dora's project page~\cite{dora}.}
    \label{fig:result_n2g}
    \vspace{-3mm}
\end{figure*}

\paragraph{Qualitative Results} We give a qualitative comparison between the generated 3D geometries of the proposed \pipelineName and other methods, as shown in Fig.~\ref{fig:result_n2g}. It impressively shows the superiority of our \pipelineName in generating high-fidelity results with rich details that are consistent with the input images, which are easily lost by other methods. Besides, our \pipelineName also produce robust generations with relatively smooth surface when less details presented in the input images (\eg the first and third example in Fig.~\ref{fig:result_n2g}). We give more generation results of our \pipelineName in Fig.~\ref{fig:more_our_results}, with more in supplementary materials.

\begin{figure}[h] \centering
    \includegraphics[width=0.42\textwidth]{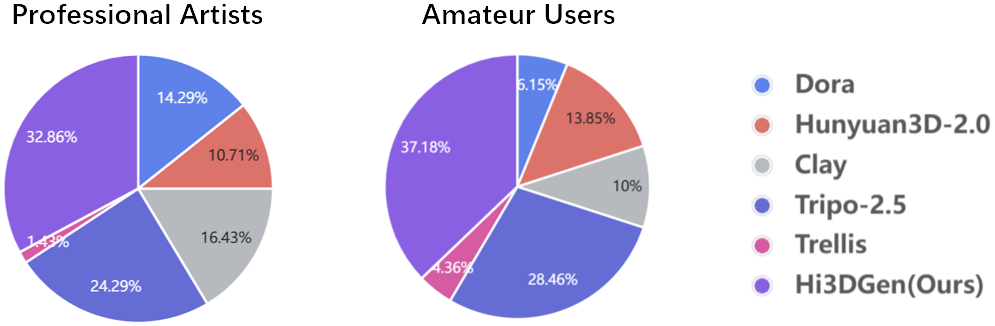}
    \caption{User study results.}
    \label{fig:user_study}
    \vspace{-3mm}
\end{figure}

\paragraph{User Study} We conducted a user study to evaluate the 3D generation results of our \pipelineName and 5 other methods including Hunyuan3D-2.0, Dora, Clay, Tripo-2.5, and Trellis. All 3D results for user study are randomly sampled from the 300$\times$6 generations for visual comparison. The evaluation criteria focus on the fidelity of the generated 3D geometry to the input images, which is measured by the consistency in both overall shape and local details. For the parts of the input images that are not visible, we ask the evaluators to exercise their judgment and imagination to assess the plausibility of the generated results and their stylistic consistency with the visible portions.
To ensure the comprehensiveness and professionalism of the user study, we invite two groups of evaluators. The first group consist of 50 amateur 3D users, who assess 100$\times$6 randomly sampled results from the perspective of everyday applications, such as 3D printing. The second group includes 10 professional 3D artists, who evaluate 20$\times$6 results from the standpoint of professional use, like 3D modeling and design. The results are presented in Fig.~\ref{fig:user_study}, which shows that our \pipelineName achieves the highest generation quality for both amateur users and professional artists.

\subsection{Ablation Study}
\label{sub:ablation}

\paragraph{Normal Bridge}
\label{parag:abla_normal}
We first validate the effectiveness of using normal maps to bridge 3D generation. A direct image-to-geometry generator based on Trellis~\cite{xiang2024trellis} performs worse than our normal-bridged Hi3DGen, and when using the same normal regularization and training data as Hi3DGen, it produces fake details (see the first two columns v.s. the last column of Fig.~\ref{fig:abl_normalbridge}). We also validate the influence of using normal conditions of different accuracy and sharpness to the final 3D generation quality. Smoother or wrong normal estimations by other methods lead to a performance drop, which also proves the importance of using accurate and sharp estimated normals as the bridge.

\begin{figure}[tb] \centering
    \includegraphics[width=0.48\textwidth, height=0.22\textwidth]{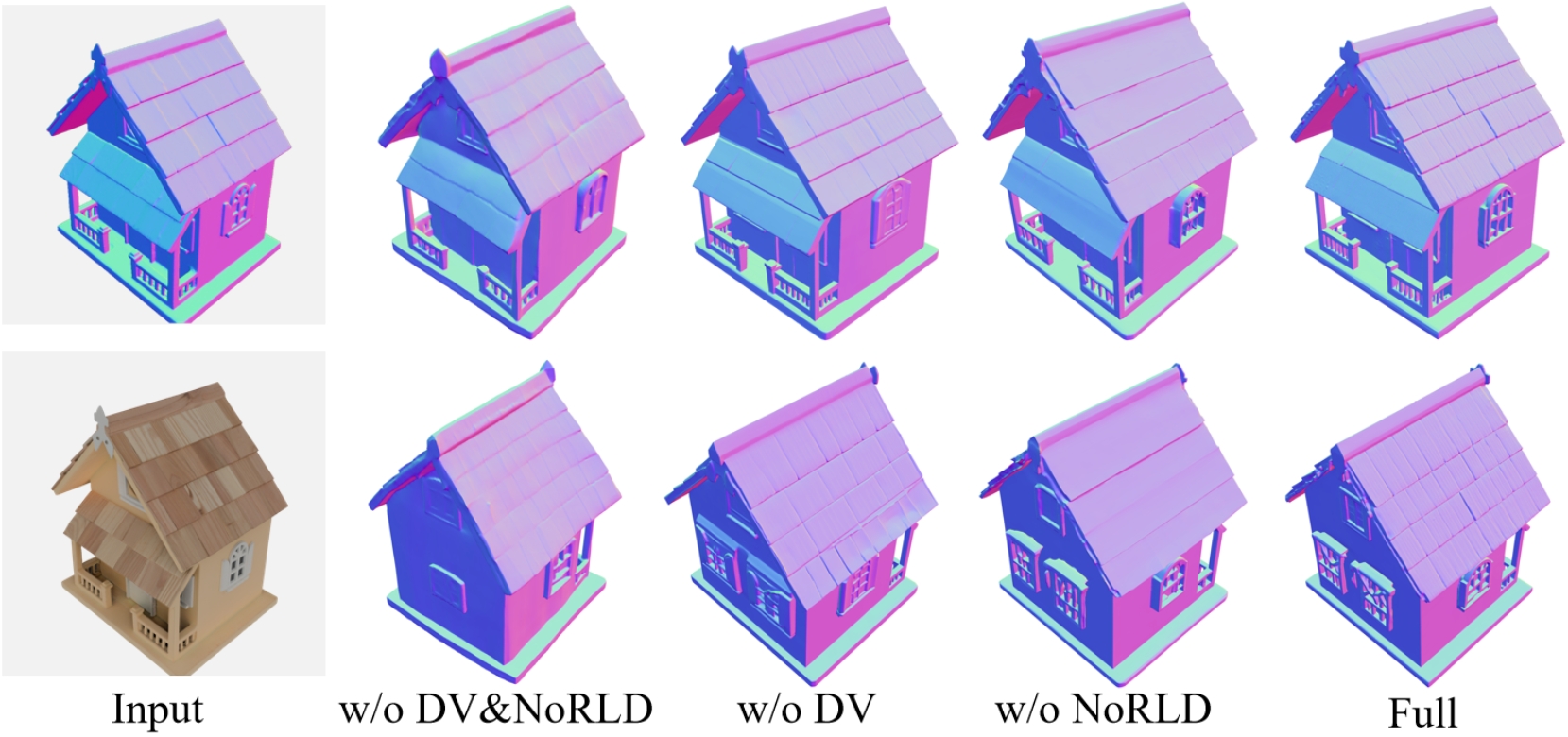}
    \caption{Ablation on the proposed \NtoGmethodName.}
    \label{fig:ablation_n2g}
    \vspace{-3mm}
\end{figure}

\paragraph{\datasetName Data}
\label{parag:abla_data}
We also validate the value of the proposed \datasetName dataset. By integrating image-normal training pairs rendered from \datasetName data, our \ItoNmethodName can achieve 0.4 and 1.7 improvements in NE and SNE respectively, as shown in the first two rows of  Tab.~\ref{tab:abl_i2n}. By using additional normal-geometry training pairs from \datasetName, our \NtoGmethodName can achieve higher-fidelity generation details, as shown in the 3rd and final columns in Fig.~\ref{fig:ablation_n2g}. 



\paragraph{\ItoNmethodName Ablation}
\label{parag:abla_operationsi2n}
We conduct ablative experiments to validate the three components in our \ItoNmethodName: the noise injection technique, the dual-stream architecture, and the domain-specific training. Results in Tab.~\ref{tab:abl_i2n} validates the effectiveness of each component. More qualitative comparisons are included in the supplementary materials. 

\paragraph{\NtoGmethodName Ablation}
\label{parag:abla_operationsn2g}
We visualize the difference of not using the proposed online normal regularization in Fig.~\ref{fig:ablation_n2g}, which shows that whether using \datasetName data for training, adopting normal regularization can greatly improve the generation fidelity (zoom in to see the roof details).

\begin{table}
    \centering
    \footnotesize
    \renewcommand{\arraystretch}{1.3}
    \caption{Ablation study on different components of \ItoNmethodName. ``DV'', ``NI",``DS", and ``DST'' denote \datasetName data, Noise Injection technique, Dual-Stream architecture, and Domain-Specific Training strategy, respectively. }
    \begin{tabular}{p{2.6cm}>
    {\centering\arraybackslash}p{1.6cm}>
    {\centering\arraybackslash}p{1.6cm}
    }
        \toprule
        \rowcolor{gray!20} \textbf{Method} & \textbf{NE} $\downarrow$ & \textbf{SNE} $\downarrow$ \\
        \midrule
        \rowcolor{gray!10} Ours (full model) & 
        \textbf{21.837} & \textbf{26.628} \\
        Ours w/o DV   & 22.209  & 28.324 \\
        Ours w/o DST  & 23.288 & 29.690 \\
        Ours w/o DS   & 21.918 & 29.520 \\
        Ours w/o all  & 22.507 & 35.997 \\
        \bottomrule
    \end{tabular}
    \vspace{-3mm}
    \label{tab:abl_i2n}
\end{table}




\section{Conclusion}
\label{sec:Conclusion}
This paper propose \pipelineName, a high-fidelity image-to-3D generation framework. It works by using normal map, a 2.5D representation, to bridge the 3D generation for rich details consistent with input images in generations. \pipelineName consists of three components, a image normal estimator producing robust, stable, and sharp predictions, a normal-to-3D synthesis network with normal regularization for geometry consistency, and a 3D data synthesis pipeline providing detail-rich synthetic data for the training. Extensive experiments validate their effectiveness and superiority. 

\paragraph{Limitations}Although \pipelineName generates detail-rich 3D results, part of them still have possible inconsistent or non-aligned details with the input, which is caused by the generative nature of the 3D latent diffusion learning. It is our future work to pursue reconstruction-level 3D generations.

\clearpage
{
    \small
    \bibliographystyle{ieeenat_fullname}
    \bibliography{ref}
}

\clearpage
\setcounter{page}{1}
\maketitlesupplementary
\renewcommand{\thetable}{S\arabic{table}} 
\renewcommand{\thefigure}{S\arabic{figure}}


\section{More Details for the Method}
\paragraph{More Implementation Details}
We reimplement GenPercept~\cite{xu2024genpercept} using StableDiffusion 2.1 by replacing the input noise with VAE-encoded images. We implement noise injection using an EDM-style noise sampler, which randomly adds noise to the encoder output latents before they are input to the decoder. Specifically, we follow EDM to use standard parameters with $\sigma_{\min} = 0.002$ and $\sigma_{\max} = 80.0$. We guarantee the SNR of the features by selecting timestamps from 0 to 400, which we empirically found maintains coarse shape knowledge — this approach aligns with Instruct-Pix2Pix, which also adds noise in a middle timerange to avoid structural changes. To better preserve coarse knowledge while instructing the encoder to focus on detailed information, we follow ControlNet~\cite{zhang2023controlnet} to add a secondary encoder by copying the weights of the SD2.1 encoder and concatenating multi-layer features with the decoder for dual-stream training. Specifically, an image is first processed by the VAE before entering the two encoders. The encoders do not share weights since they are designed to learn different frequency information from the images.

\paragraph{Training Details} For training our I2N method, we input identical image latents to the dual encoders, where the coarse encoder remains noise-free with no modifications to any layers, while for the fine-grained encoder, we follow the approach described above to inject noise on the encoder output. Notably, we feed the noised features to the decoder layers rather than back to the encoder layers. For domain-specific training, we first train on real-world data from the Depth-pro dataset for 50,000 steps with a batch size of 256 at 768px resolution. We randomly crop images at varying aspect ratios before resizing to 768px. Subsequently, we train our model on rendered images from DetailVerse and Objaverse\cite{deitke2023objaverse}. For DetailVerse, we render 40 spherical views per object using nvdiffrast, varying the radius and field of view. For Objaverse training, we use the 40-view renders from GObjaverse, applying the filter criteria from RichDreamer\cite{qiu2024richdreamer} to select 170K high-quality samples. We fine-tune in this second stage on the synthetic dataset while freezing the coarse encoder. For training our N2G method, we follow Trellis to employ rectified flow for model fine-tuning. We reuse the Sparse Structure VAE and Structured Latent VAE without modification, as DetailVerse is generated using Trellis (ensuring domain compatibility), and our selected Objaverse subset is already included in the original Trellis training dataset.

\begin{figure}[tb] \centering
    \includegraphics[width=0.48\textwidth, height=0.24\textwidth]{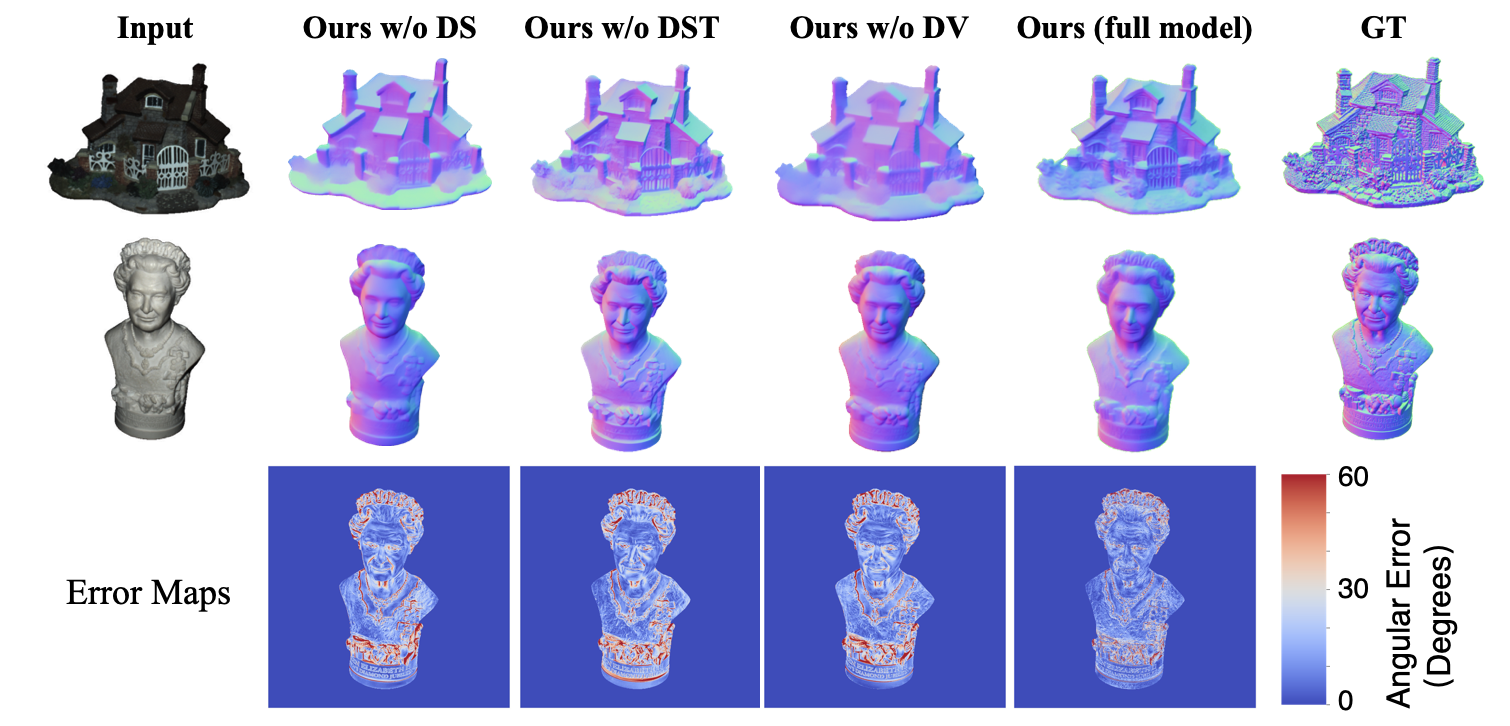}
    \caption{Ablations on image-to-normal estimation. }
    \label{fig:abl_i2n}
    \vspace{-3mm}
\end{figure}

\begin{figure}[tb] \centering
    \includegraphics[width=0.35\textwidth, height=0.26\textwidth]{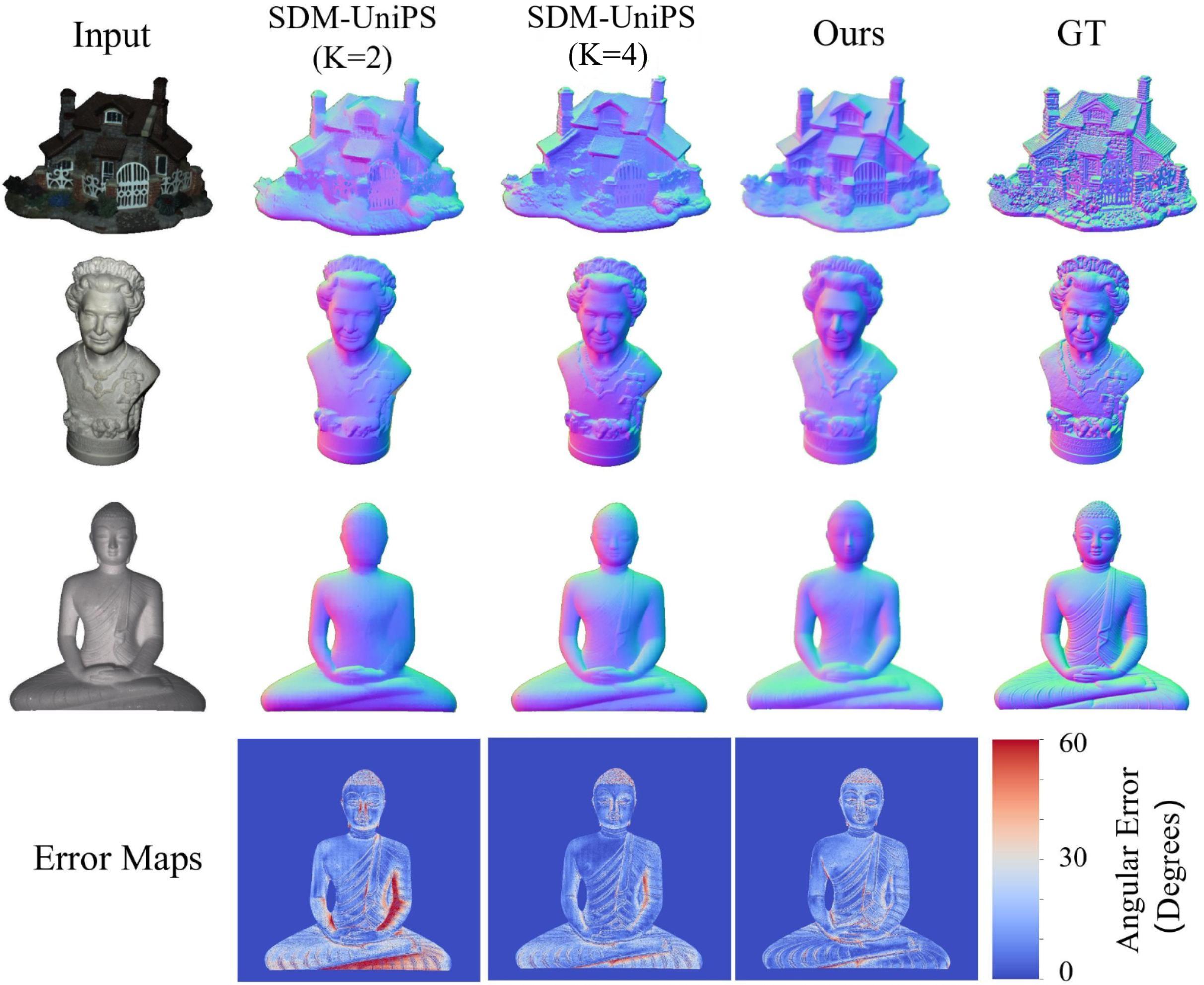}
    \caption{Qualitative comparison of image-to-normal estimation with SOTA Photometric Stereo-based Method, SDM-UniPS.}
    \label{fig:method_unips}
    \vspace{-3mm}
\end{figure}

\begin{table*}[h]
    \centering
    \footnotesize
    \setlength{\tabcolsep}{5pt} 
    \renewcommand{\arraystretch}{1.2} 
    \caption{Image-to-Normal estimation evaluation on Luces-MV (SNE). Comparisons of NiNRE with SOTA photometric stereo techniques. \textbf{Bold} indicates the second best results and \textcolor{red}{\textbf{Red}} indicates best results.}
    \begin{tabular}{lccccccccccc} 
         \toprule
         Method & Bowl & Buddha & Bunny & Cup & Die & Hippo & House & Owl & Queen & Squirrel & Ave. \\
         \midrule
         SDM-UniPS (K=2) & 37.65 & 26.24 & \textbf{29.02} & 23.70 & \textbf{26.32} & 31.45 & 40.68 & \textbf{24.56} & 27.14 & 26.10 & 29.286 \\
         SDM-UniPS (K=4) & \textcolor{red}{\textbf{31.64}} & \textcolor{red}{\textbf{20.59}} & \textcolor{red}{\textbf{23.23}} & \textbf{23.39} & \textcolor{red}{\textbf{25.58}} & \textcolor{red}{\textbf{21.91}} & \textbf{38.61} & \textcolor{red}{\textbf{22.26}} & \textcolor{red}{\textbf{25.97}} & \textcolor{red}{\textbf{24.04}} & \textcolor{red}{\textbf{25.722}} \\
         \rowcolor{gray!10} 
         \textbf{Ours} & \textbf{34.55} & \textbf{21.13} & 30.45 & \textcolor{red}{\textbf{17.47}} & 27.20 & \textbf{24.64} & \textcolor{red}{\textbf{34.58}} & 25.15 & \textbf{26.82} & \textbf{24.29} & \textbf{26.628} \\
         \bottomrule
    \end{tabular}
    \vspace{-3mm}
    \label{tab:method_unips}
\end{table*}


\paragraph{Inference Details}
During inference of Hi3DGen, we first utilize an off-the-shelf background removal model to isolate the foreground object. We crop the foreground and pad the image to a square format, then resize it to 768$\times$768 resolution before inputting it to our \ItoNmethodName model. During the inference of \ItoNmethodName, we do not inject any noise into the encoder features to ensure stable inference and maximize the preservation of detail information captured by the fine-grained encoder. Given the estimated normal map, we set the background to white and input the normal map to \NtoGmethodName. Trellis employs a two-stage generation pipeline to produce structured latents, which first generates the sparse structure, followed by the local latents attached to it. Following the same approach as Trellis, we first generate the sparse structure represented by sparse voxels, then initialize noise on the sparse voxel representation to generate the final structure latents using our fine-tuned structure latents flow model. The final mesh is generated using the pre-trained mesh decoder.

\paragraph{Metric Explanation} For comprehensive evaluation of image-to-normal, we adopt metrics from Dora~\cite{chen2024dora} to quantify normal map accuracy, with particular emphasis on sharp edges where geometric details are most salient. Specifically, we compute the Sharp Normal Error (SNE) through a three-step process: Firstly, we detect salient regions in the ground truth normal maps through canny. Secondly, we dilate these masked regions to ensure complete coverage of edge features. Finally, we calculate the normal angle error within these masked regions. For completeness and fair comparison with existing methods, we also report the Normal Error (NE) across the entire normal map, measured in degrees.  For evaluating normal-to-geometry conversion, we render normal maps from 22 fixed, evenly spaced viewpoints around each object using nvdiffrast~\cite{Laine2020diffrast}, which is used to compute SNE and NE.

\section{More Details for the DetailVerse}
To ensure the quality of our synthesized meshes, we implement a rigorous multi-stage data generation and filtering pipeline that combines expert evaluation with automated assessment techniques.

\paragraph{Step 1: Semantic Text Prompt Curation} 
We initiate the 3D data synthesis process with text prompts rather than image prompts, as textual descriptions enable more precise control over semantic diversity, thereby ensuring variety in the resulting geometries.
To collect high-quality text prompts with semantic diversity, we first sourced approximately 14M raw prompts from DiffusionDB~\cite{wang2023diffusiondb}, covering a wide range of topics relevant to AI generation applications. We employed a LLaMA-3-8B model~\cite{touvron2023llama}, fine-tuned with manually annotated examples, to categorize these prompts into four distinct classes: (i) Single Objects; (ii) Multiple Objects; (iii) Scenes; and (iv) Others. Only prompts from classes (i) and (ii) were retained, yielding approximately 1M high-fidelity prompt candidates.

Next, we applied rule-based filtering to preserve geometric and semantic attributes while eliminating stylistic modifiers. Empirically, we observed that input images with near-isometric viewpoints and CGI-rendered aesthetics significantly enhance the fidelity of 3D synthesis. Thus, we implemented structural prompt standardization to prompting the image generation. Specifically, we applying domain-specific prompt templates to enforce explicit geometric cues and structural clarity (e.g., ``isometric perspective'', ``Unreal Engine 5 Rendering'', ``4K'', ``MasterPiece'').  This comprehensive process yielded approximately 1.5 million well-curated and natural prompts.

\paragraph{Step 2: High-Quality Image Generation} 
With our diverse text prompt collection established, the next step involved generating corresponding images suitable for 3D asset synthesis. The key requirements for these images were: (i) high visual fidelity with rich details that accurately reflect the textual descriptions; and (ii) specific viewpoints and styles that facilitate robust 3D reconstruction.

We integrated the state-of-the-art Flux.1-Dev~\cite{flux} as our image generator. To ensure detailed output, we filtered the generated images by ranking their sharpness according to the number of sharp pixels, as calculated using Canny edge detection, and retained only the top 50\%. For each prompt, we randomly selected a seed to encourage variety, generating exactly one image per prompt.

To mitigate geometry distortion in the resulting 3D models, we utilized OrientAnything~\cite{wang2024orient}, a robust object orientation estimation model, to measure the alignment between the camera view and canonical object orientation. Images with angular deviations exceeding $60^\circ$ were rejected to prevent structural distortions and preserve geometric fidelity. Through this filtering process, we preserved 1 million high-quality images for the subsequent 3D synthesis stage.

\paragraph{Step 3: Robust Image-to-3D Synthesis} 
We employed Trellis~\cite{xiang2024trellis}, a state-of-the-art two-stage 3D generator, to produce high-fidelity 3D objects from the prepared images. Given its superior performance with high-quality inputs, we initially generated a set of preliminary meshes.

To ensure mesh quality, we implemented a rigorous data cleaning process combining expert evaluation with automated assessment. We randomly sampled 10K meshes and engaged 10 trained experts to conduct triple-blind quality assessments. The evaluation criteria primarily focused on surface quality, specifically examining whether the rendered normal maps contained holes or noise artifacts.

\begin{figure*}[tb] \centering
    \includegraphics[width=\textwidth]{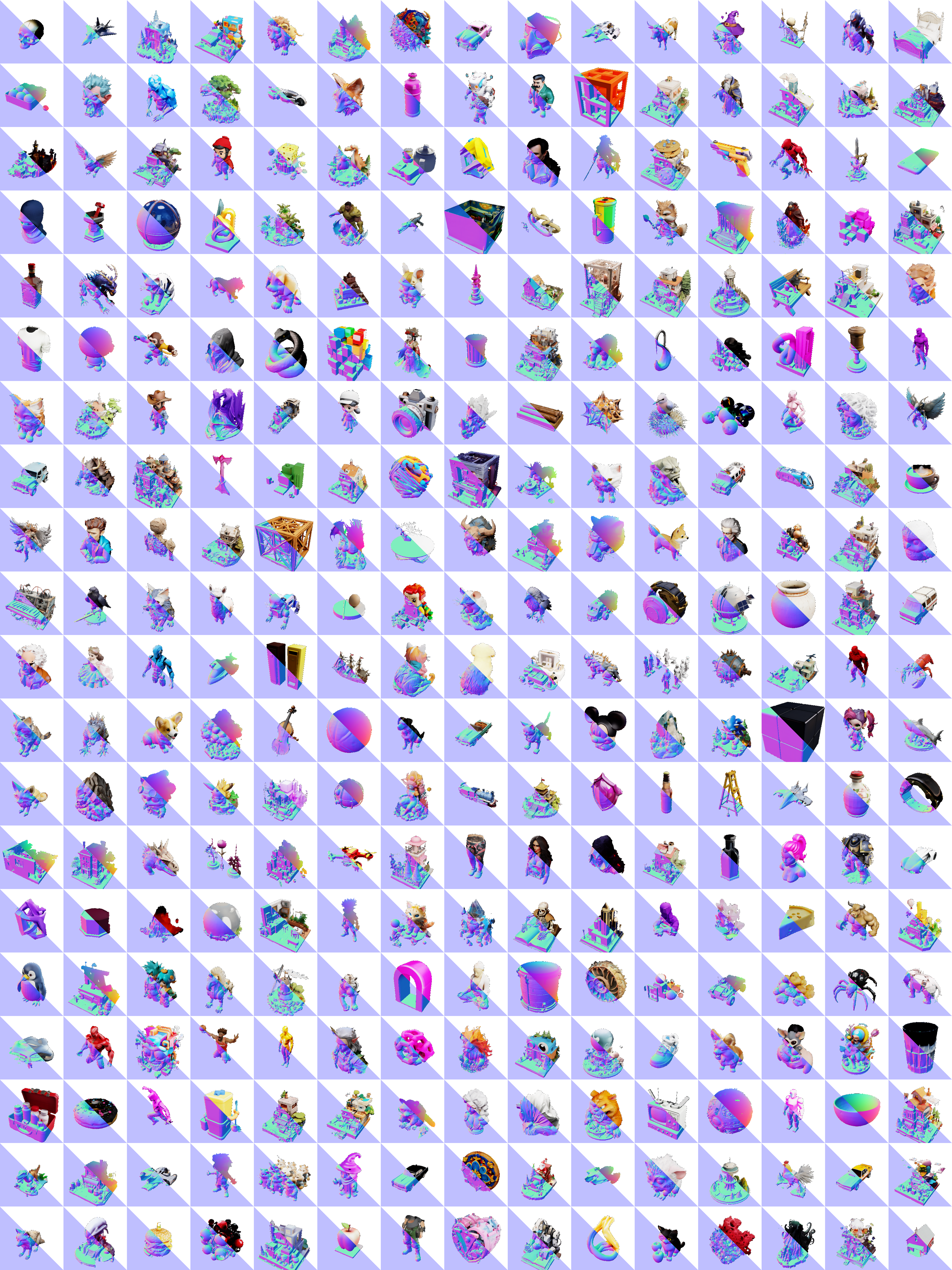}
    \caption{More \datasetName data exhibition. }
    \label{fig:more_dv_data}
    \vspace{-3mm}
\end{figure*}

Based on these expert annotations, we trained a quality assessment network using DINOv2~\cite{oquab2024dinov2} features. Specifically, we extracted features from four equiangular rendered normal maps of each mesh and trained a three-layer MLP classifier for quality scoring. This trained network was then applied to evaluate the entire dataset. Models that received positive classifications across all four views were selected for training our \NtoGmethodName model. Through this comprehensive quality assurance process, we retained 700K high-quality object meshes to form our \datasetName dataset. A data gallery is shown in Fig.~\ref{fig:more_dv_data}, and better visualizations are presented in the demo video.

\section{More Ablation Studies}
\paragraph{\ItoNmethodName Ablation}
We provide qualitative results to supplement the ablation studies on the proposed \ItoNmethodName. As shown in Fig.~\ref{fig:abl_i2n}, each component makes positive role in the final performance. 


\section{More Results}
\begin{figure*}[tb] \centering
    \includegraphics[width=\textwidth]{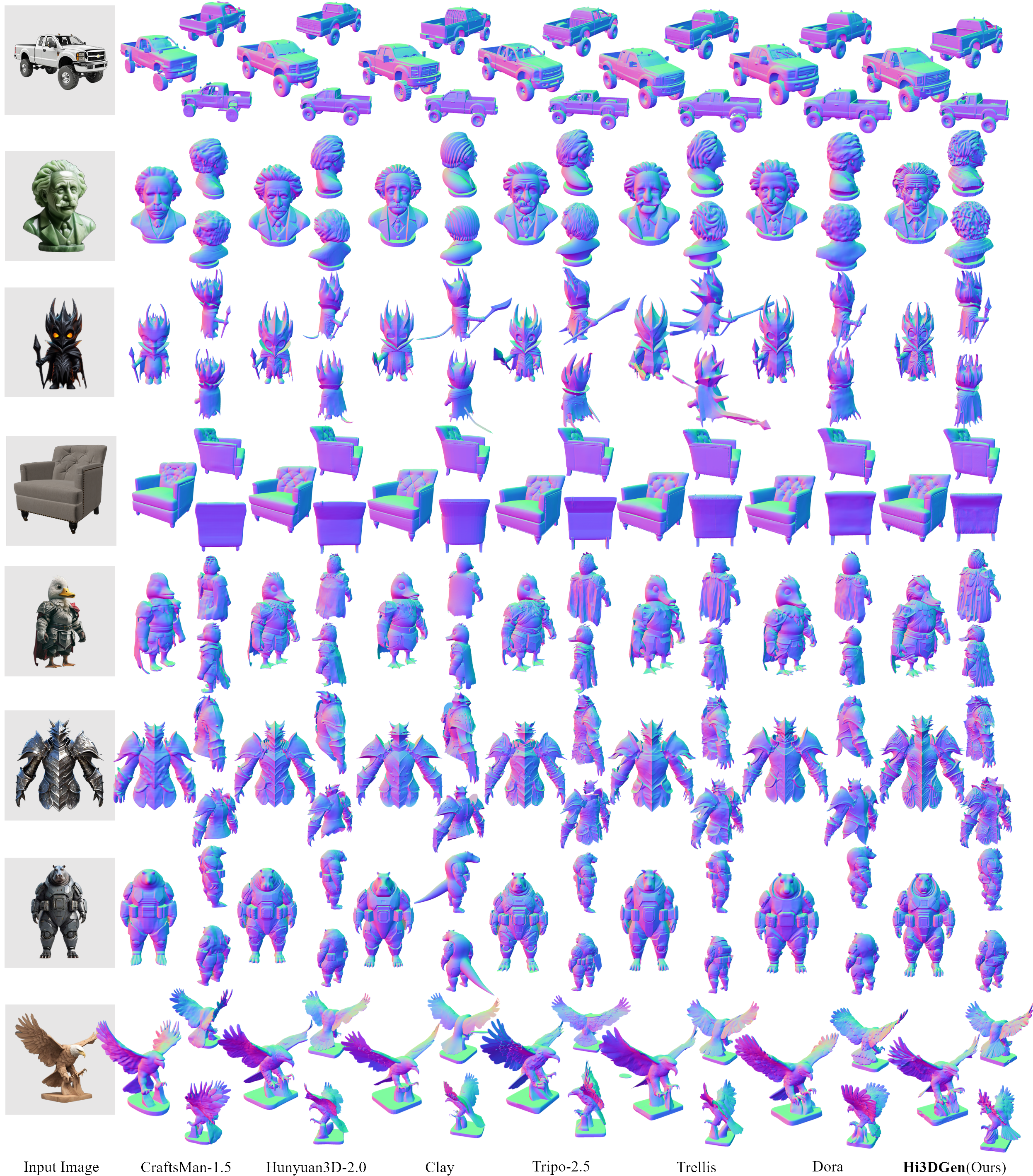}
    \caption{More 3D generation results comparison. }
    \label{fig:more_comparison}
    \vspace{-3mm}
\end{figure*}

\paragraph{More Image-to-Normal Results} We compare \ItoNmethodName with SOTA photometric stereo technique (SDM-UniPS~\cite{ikehata2023sdm-unips}), which works in a different setup that requires input images under $K$ different lightning conditions. (As shown in Fig. ~\ref{fig:method_unips}). 
\paragraph{More Comparisons} We give more qualitative comparisons in Fig.~\ref{fig:more_comparison}, which shows our normal-bridged \pipelineName can achieve more consistent 3D detailed geometries with input images than existing methods. Better visualizations are presented in the demo video.

\end{document}